\documentclass{achemso}
\setkeys{acs}{usetitle=true}
\usepackage{amsmath}
\usepackage{bm}
\usepackage{amsfonts}
\usepackage{braket}
\usepackage{color}
\usepackage{graphicx}
\usepackage{braket}

\title{Quantum embedding theories}
\author{Qiming Sun}
\author{Garnet Kin-Lic Chan}
\affiliation{Division of Chemistry and Chemical Engineering, California Institute of Technology,
  Pasadena, California 91125, USA}
\email {gkc1000@gmail.com}

\begin{document}

\maketitle

\section{Conspectus}

In complex systems, it is often the case that the region of interest forms only
  one part of a much larger system. The idea of joining two different quantum simulations -
  a high level calculation on the active region of interest, and a low level calculation
  on its environment - formally defines a {\it quantum embedding}.
  While any combination of techniques constitutes an embedding,
  several rigorous formalisms have emerged that provide for exact
  feedback between the embedded system and its environment. These three formulations: {\it density functional embedding},
  {\it Green's function embedding}, and {\it density matrix embedding}, respectively use the single-particle density,
  single-particle Green's function, and single-particle density matrix as the quantum variables of interest.
  
  Many excellent reviews exist covering these methods individually. However, a unified presentation
  of the different formalisms is so far lacking. Indeed, the various languages commonly used:
  functional equations for density functional embedding; diagrammatics for Green's function embedding; and
  entanglement arguments for density matrix embedding, make the three formulations appear vastly different.
  In this account, we introduce the basic equations of all three formulations
  in such a way as to highlight their many common intellectual strands. 
While we focus primarily on a straightforward theoretical perspective,
  we also give a brief overview of recent applications, and possible future developments.

  The first section starts with density functional embedding, where we introduce the key embedding potential via the
   Euler equation.
  We then  discuss recent work concerning the treatment of the non-additive kinetic potential, 
  before describing mean-field density functional embedding, and wavefunction
  in density functional embedding. 
  We finish the section with extensions to time-dependence and
  excited states. 

  The second section is devoted to Green's function embedding. Here, 
  we use the Dyson equation to obtain equations that parallel as closely as possible
  the density functional embedding equations, with the hybridization playing the role
  of the embedding potential. 
  Embedding a high-level self-energy
  within a low-level self-energy is treated analogously to wavefunction in density functional
  embedding. The numerical computation of the high-level self energy allows
  us to briefly introduce the bath representation in the quantum impurity problem. 
  We then consider translationally invariant systems to bring in the important
  dynamical mean-field theory. Recent developments to incorporate
  screening and long-range interactions are discussed. 

  The third section concerns density matrix embedding. Here, we first highlight some mathematical complications 
  associated with a simple Euler equation derivation, arising
  from the open nature of fragments. This motivates the density matrix
embedding theory, where we use the Schmidt decomposition to represent the entanglement
through bath orbitals.
The resulting impurity plus bath formulation resembles that of dynamical mean-field theory. We discuss
the numerical self-consistency associated with using a high-level correlated wavefunction with a mean-field low-level treatment,
and connect the resulting numerical inversion to that used in density functional embedding.  

We finish with  perspectives   on the future of all three methods.

\section{Introduction}

Embedding theories provide a natural way to focus computation on a small region
within a larger environment, such as atoms near the active site in an enzyme.
In any embedding theory, the full problem is partitioned into the fragment of interest,
($A$), and its environment ($B$). (The terms ``subsystem''  or ``impurity''
are commonly used in place of ``fragment''). 
There may be multiple
fragments of interest, and the environment may also be partitioned.
In any case, the purpose of the embedding
is to provide a computational recipe for the properties of $A$, taking into account
its environment, without the computational cost of treating the full problem.
The idea is thus very general, and  encompasses a wide variety of methods. For example,
whenever $A$ and $B$ are treated at two different levels, e.g. when
freezing orbitals in a calculation, we are formally performing embedding!
What concretely distinguishes one embedding theory from another, is the particular
way in which the effects of the environment are communicated to the fragment $A$, and vice-versa. 

Here, we will describe a family of three related, rigorous, quantum embedding theories:
density functional (DFT) embedding\cite{Cortona1991,Wesolowski1993,Jacob2014,libisch2014embedded,wesolowski2015frozen},
Green's function embedding\cite{inglesfield1981method,Inglesfield2015,georges1992hubbard,Kotliar2006,Chan2011a,Chibani2016}
and density matrix embedding\cite{Knizia2012,Knizia2013,Wouters2016,Pernal2016}. Compared
to simpler embedding techniques, these stand out as they 
provide for non-trivial communication between the fragment and environment.
In the three theories, information is communicated via the density,
the single-particle Green's function, and
the single-particle density matrix, respectively. Note that we identify the kind of embedding
by its functional dependence on a quantum variable, rather than  intermediary computational objects. 
Thus, even if a density matrix is used in the calculation,
if it ultimately encodes a density functional (e.g. in a Kohn-Sham theory), then the embedding is a density functional embedding. 
There are excellent reviews for each of the three techniques individually, for example, Refs.~\cite{libisch2014embedded,Jacob2014,wesolowski2015frozen,Inglesfield2015,Kotliar2006,Wouters2016}.
Here, we will give an introduction to all three together,
that emphasizes their common intellectual strands, and provide a summary of their strengths and weaknesses.

\section{DFT embedding}

We will begin with the simplest quantum embedding, DFT embedding. 
We provide a particular perspective to make connections
to the subsequent Green's function and density matrix embedding formalisms; other perspectives 
can be found in the literature~\cite{libisch2014embedded,Jacob2014,wesolowski2015frozen}.
The formal groundwork was developed by Cortona~\cite{Cortona1991} and Wesolowski and Warshel\cite{Wesolowski1993}. 
In the ground-state version, the 
density of $A$, $\rho_A$, is adjusted through an 
external field, $v_A$, and we view this field as 
coming from the environment. Further, the energy of $A$ is modified by its interactions
with the environment, through the Coulomb term, and indirectly
through the exclusion principle. These energetic contributions are 
contained in a term $\Delta E$. 

The DFT embedding formalism provides rigorous foundations for $v_A$ and $\Delta E$ as
density functionals.
We begin with the exact energy density functional for the full system,  $E[\rho]$, which
 determines the ground-state density through the Euler equation
\begin{align}
  \frac{\delta E[\rho]}{\delta \rho} - \mu = 0,
\end{align}
where $\mu$ is the chemical potential. 
Partitioning the energy into its  Kohn-Sham kinetic, Coulomb, external, and 
exchange-correlation pieces~\cite{parryang}, 
\begin{align}
E[\rho] = T_s[\rho] + J[\rho] + V_{ext}[\rho] + E_{xc}[\rho],
\end{align}
we rewrite the Euler equation 
\begin{align}
  \frac{\delta T_s[\rho]}{\delta \rho} + v_J[\rho] + v_{ext} + v_{xc}[\rho] - \mu = 0. \label{eq:euler_pieces}
\end{align}
$E_{xc}[\rho]$ and $v_{xc}[\rho]$ account for the non-trivial correlation effects, and
must be approximated. When necessary, we indicate the dependence of
these quantities on the approximation scheme $S$ by writing them as
$E_{xc}^S[\rho]$, $v_{xc}^S[\rho]$.

We now derive the expressions for $v_A$ and $\Delta E$. 
Splitting $E[\rho]$ into the fragment energy and its remainder, $E[\rho]=E[\rho_A]+\Delta E[\rho,\rho_A]$ ($E_A = E[\rho_A]$),
stationarity with respect to $\rho_A$ implies
\begin{align}
\frac{\delta}{\delta \rho_A} \Bigl[E[\rho_A] + \Delta E[\rho,\rho_A]\Bigl] - \mu = 0. \label{eq:dft_stationary}
\end{align}
This is the same as the Euler equation for $A$ placed in an external
field $v_A$, if we choose 
\begin{align}
v_A= \frac{\delta \Delta E[\rho,\rho_A]}{\delta \rho_A}  \label{eq:exact_dft_v}.
\end{align}
Eq.~(5) thus defines the exact embedding potential, which yields the exact
subsystem density $\rho_A$. 
Note we use $\rho$, $\rho_A$ as the
working variables. The use of a global and a fragment quantum
variable is a choice common to all three embedding formalisms in this article. However,
in DFT embedding, one often works instead with the subsystem and environment densities separately, $\rho_A, \rho_B$,
with $\rho = \rho_A + \rho_B$. This has advantages in practice, for example,
if $\rho_A$ and $\rho_B$ are ensemble $N$-representable, then so is $\rho$~\cite{wesolowski2008embedding}. 

We express $v_A$ in terms of its components as
\begin{align}
v_A &= \frac{\delta}{\delta \rho_A}\Bigl[ T_s[\rho]-T_s[\rho_A]\Bigl] + v_J[\rho-\rho_A]
+ \frac{\delta}{\delta \rho_A}\Bigl[E_{xc}[\rho]-E_{xc}[\rho_A]\Bigl]\notag\\
&= v_{s}^\Delta + v_J^\Delta + v_{xc}^\Delta \label{eq:dft_partition}
\end{align}
where $v_{s}^\Delta$ is the {\it non-additive kinetic potential}, $v_J^\Delta$
is the environment Coulomb  potential, and $v_{xc}^\Delta$ is
the non-additive exchange-correlation potential. $v_{s}^\Delta$
is the largest contribution and expresses the exclusion principle which forces electrons in the fragment
 to occupy states orthogonal to those in the environment~\cite{Cortona1991,Wesolowski1993}. 

Commonly in DFT calculations, $E_{xc}[\rho]$ is approximated as an
 explicit functional of $\rho$ and its derivatives, (e.g. as is in the LDA or GGA's).
Consequently, $v_{xc}^{\Delta}[\rho]$ can be obtained by 
analytical functional differentiation of $E_{xc}[\rho] - E_{xc}[\rho_A]$. $T_s[\rho]$, however, is only known as
an {\it implicit} density functional through the Kohn-Sham orbitals. One way to evaluate
 $v_{s}^\Delta$ is to approximate $T_s$ by an
 explicit density functional approximation, such as
 the Thomas-Fermi (or a related) functional,
  from which the kinetic potential can be directly derived. 
This approximation was widely used
in the early days of DFT embedding~\cite{Cortona1991,Wesolowski1993}, but is
limited by the accuracy of the approximate kinetic energy~\cite{Wesolowski1996,Visscher2009}. It has found most
success in applications where the fragment is weakly bound
to the environment, for example in van der Waals complexes~\cite{Wesolowski2007},
highly ionic crystals~\cite{Cortona1992}, and in  solvation~\cite{Neugebauer2005,Wesolowski2010}. 

A more recent strategy has been to compute $\delta T_s/\delta \rho_A$ 
and the non-additive kinetic potential numerically\cite{roncero2008inversion,fux2010accurate,Goodpaster2010,Chen2011}.
Since the forward computation of $v_s \to \rho$ (to determine the non-interacting ground-state
density in an external potential $v_s$) is cheap ($O(N^3)$ where $N$ is the number of electrons) the inversion
is tractable~\cite{Zhao1992,Wu2003}. 
In practice, there are numerical difficulties~\cite{Scuseria2006}, because
large changes in $v_s$ can yield only small changes in $\rho$.
This leads to unphysical oscillations in the numerically determined $v_s$~\cite{Scuseria2006,Yang2007}. 

The above is a basic formulation of ground-state DFT embedding, and extensions exist,
for example, to ensemble representations for fragments with non-integer particle number~\cite{pernal2009orbital,elliott2009density,pfet},
 and Kohn-Sham projector formalisms to avoid the non-additive kinetic potential altogether~\cite{Manby2012,tamukong2014density}.
However, we now consider the two common contexts in which DFT embedding is applied. The
first is DFT in DFT embedding~\cite{Wesolowski1993},
where both the subsystem $A$ and environment $B$ are treated with DFT. This may not seem to result
in any simplification, but there are several ways to obtain savings.
For example, different levels of DFT can be used for $A$ and $B$ (e.g. functionals with
or without exact exchange)~\cite{fornace2015embedded} or the self-consistency may be approximated (e.g.
by replacing the self-consistent $\rho_B$ by a frozen superposition of densities). The latter is particularly
appropriate for solvent systems~\cite{Neugebauer2005,Wesolowski2010}, 
and allows for very large environments, such as protein frameworks, to be considered.

The second common context is where the fragment
is described using a high-level wavefunction, and the environment
by DFT, and embedding  connects the two~\cite{Carter1998,wesolowski2008embedding,Hofener2012,Goodpaster2012,Visscher2013b}.
Wavefunction in DFT embedding was pioneered by Carter and
coworkers~\cite{Carter1998,Chen2011} and has
attracted much recent attention, because of the difficulties
of DFT in treating aspects of electronic structure such as excited states and bond-breaking.

In wavefunction in DFT embedding, two approximations enter for the correlations, $S_{WF}$ and $S_{DFT}$.
The energy functional is defined as
\begin{align}
  E[\rho] = E^{S_{WF}}[\rho_A] + E^{S_{DFT}}[\rho] - E^{ S_{DFT}}[\rho_A] \label{eq:totalenergyE}
  \end{align}
where the wavefunction energy for fragment $A$ is formally expressed through
\begin{align}
E^{S_{WF}}[\rho_A] = \min_{\Psi^{S_{WF}} \to \rho_A} \langle \Psi^{S_{WF}}|\hat{H}_A|\Psi^{S_{WF}}\rangle. \label{eq:constrained_dft_wf}
\end{align}
Since the effect of the environment is completely contained within  $v_A$, 
{the minimizing $\Psi$ in Eq.~(8) is the ground-state of $\hat{H}_A + v_A$}. Any wavefunction ansatz
may be used to approximate this eigenstate~\cite{Carter1998,Chen2011,Hofener2012,Goodpaster2012,Visscher2013b}, and existing quantum chemistry programs only need be modified to include the potential $v_A$, which adds to the attractiveness of the method.

In the simplest scheme, $\rho_A$ and $v_A$ are defined both from the DFT density,
and the DFT expression, $\Delta E^{S_{DFT}}[\rho, \rho_A]$ in Eq.~(5).
  In a more sophisticated scheme, the  wavefunction treatment of correlation is used
  to improve the density self-consistently. This is achieved by defining the exchange-correlation
potential of the full system 
\begin{align}
  v_{xc}[\rho] = v_{xc}^{S_{WF}}[\rho_A] + v_{xc}^{S_{DFT}}[\rho] - v_{xc}^{S_{DFT}}[\rho_A ]. \label{eq:vxc_full}
\end{align}
Self-consistent iteration between $v_{xc}$  and $v_A$ 
then allows for the DFT density $\rho_A^{DFT}$ and the WF density $\rho_A^{WF}$ to become identical.

There is an increasing number of applications using wavefunction in DFT embedding, such as
to molecules adsorbed on 
surfaces~\cite{Carter2001,Carter2012a} and molecular fragments embedded in larger systems~\cite{goodpaster2014accurate}.
A growing community is exploring these techniques also for
excited state properties (see e.g. Ref.~\cite{SeveroPereiraGomes2012}).
The simplest way to compute excited states is to assume
that the ground-state and excited-state embedding potential and energy functional
are identical\cite{Carter2002}. This is, however, an approximation,
and in principle, the ground-state formalism must be extended.
There are two ways to do so.
The first is a state-specific DFT
embedding, where $E[\rho]$ and $v_A$ acquire an excited state dependence.
State-specific DFT embedding has been explored by several workers~\cite{Neugebauer2013a,prager2016first},
who find that  state-dependence  gives significant corrections, especially
when the charge character of the excited state differs from the ground-state. 
A second way is through time-dependent DFT, where excitation energies are poles of the response. Here,
the embedding potential becomes time-dependent, with a non-local dependence on the density at
earlier times, $v_A[\rho(t')]$~\cite{casida2004generalization,Neugebauer2007,Neugebauer2009a,Neugebauer2013,pavanello2013subsystem}. Applications using this second formalism are now beginning to appear~\cite{chulhai2016external}.

Despite its conceptual simplicity, wavefunction in DFT embedding inherits limitations
intrinsic to all combinations of wavefunctions with density functional approximations. For example,
in the total energy expression, Eq.~(7), the
non-additive part of the energy is described at the 
DFT level, while an accurate total energy requires error cancellation between the wavefunction
and DFT descriptions of the system ($E^{S_{WF}}[\rho_A] - E^{S_{DFT}}[\rho_A]$). Incomplete cancellation
is sometimes referred to as ``double counting''. The two sources of error are important if the
interface between the fragment and environment cuts across a bond of interest, as they affect
the correlation energy of the bond.
Similarly, if van der Waals' interactions between the fragment and environment are important,
such contributions, omitted in many density functionals, will be missed. 
Both these situations can be remedied formally by increasing the system size treated
with the wavefunction method, albeit at increased cost.

Another disadvantage of DFT embedding is that it is difficult, through examining $\rho_A$
alone, to distinguish between a
fragment which is bonded with the environment, 
and one which is not.
This is because the density, by definition, does not contain direct information 
on off-diagonal density matrix correlations (i.e. coherence and entanglement). While, in principle, all effects can be captured by the exact density functional,
the lack of the off-diagonal information can pose difficulties
for density functional approximations used in practice. One way to surmount this is to consider
embedding theories of richer quantum variables with off-diagonal correlations, such as the single-particle Green's function 
or the single-particle density matrix. We now turn to these embedding formalisms.

\section{Green's function embedding}

Green's function embedding has a long history, and is widely used in condensed matter problems~\cite{grimley1974chemisorption,inglesfield1981method,georges1992hubbard,Kotliar2006,Inglesfield2015,Chibani2016}.
We cannot cover the extensive literature on
models, and will restrict ourselves to
the aspects of Green's function embedding relevant to ab-initio quantum chemistry~\cite{Chan2011a,Lin2011,Chibani2016,Turkowski2012,Lan2015}.

The zero-temperature, time-ordered, single-particle Green's function generalizes
the familiar single-particle density matrix, to carry additional information on time-dependent correlations. 
In a basis labelled by $i,j$, it is defined as
\begin{align}
i G_{ij}(t) = \mathcal{T} \langle \Psi_0 | a_i (0) a^\dag_j(t) | \Psi_0\rangle
\end{align}
where $\mathcal{T}$ denotes time-ordering, $\Psi_0$ the ground-state wavefunction, and $a^{(\dag)}$ are
electron annihilation (creation) operators in the Heisenberg representation. It is common also to use
$\mathbf{G}(\omega)$, the Fourier transform of $\mathbf{G}(t)$. The imaginary
part of the Green's function $-\frac{1}{\pi} \mathrm{tr} \ \mathrm{Im} \ \mathbf{G}(\omega)$
is the single-particle density of states, while the equal-time Green's function $-i \mathbf{G}(t=0_+)$ (where $0_+$ is a positive infinitesimal) is the single-particle density matrix. As an example, the real-space non-interacting Green's function is given by $\mathbf{g}(\mathbf{r}, \mathbf{r'}, \omega) = \sum_i \phi_i^*(\mathbf{r}) \phi_i (\mathbf{r'}) (\omega  - \epsilon_i + i0_+)^{-1}$, where $\phi_i, \epsilon_i$
are the single-particle orbitals and energies.

Unlike in DFT, the  energy is computable explicitly from the exact Green's function as
\begin{align}
  E=\frac{1}{2} \int_{-\infty}^{\mu} d\omega \ \mathrm{tr} \ \Bigl[ (\mathbf{h} + \omega \mathbf{1}) \ \mathrm{Im} \ \mathbf{G}(\omega)\Bigl] \label{eq:midgal}
\end{align}
($\mu$ is the chemical potential, $\mathbf{h}$ is the single-particle Hamiltonian).
For approximate Green's functions, the expression must be modified with additional terms to obtain a
variational bound~\cite{almbladh1999variational}. One example is the Luttinger-Ward functional, from which an
embedding formalism can be constructed\cite{potthoff2003self}.
However, it is complicated to specify and we do not need all its properties here.
Instead, it is sufficient to consider the corresponding Euler equation,
namely the Dyson equation. 

The Dyson equation relates the Green's function of different Hamiltonians. For example,
the Green's function of a non-interacting system $\mathbf{g}(\omega)$ and that of 
 the interacting system $\mathbf{G}(\omega)$ are related by
\begin{align}
\mathbf{G}(\omega) = \mathbf{g}(\omega) + \mathbf{G}(\omega) \boldsymbol{\Sigma}(\omega) \mathbf{g}(\omega)
\label{eq:dyson}
\end{align}
where $\boldsymbol{\Sigma}(\omega)$, the self-energy, accounts for interactions.
Eq.~(12) is the analog of the DFT Euler equation Eq.~(3)
where $\mathbf{g}(\omega)$ plays a similar role to $\delta T_s/\delta \rho + v_{ext}$, and the self-energy plays the
part of the Coulomb plus exchange-correlation potential.
In practice, the self-energy must be approximated, and it is convenient
to discuss such approximations in diagrammatic terms. The exact (proper) $\boldsymbol{\Sigma}(\omega)$ 
is a sum over all perturbation theory diagrams, where each diagram is a connected graph of Green's function 
and interaction lines, and no diagram can be cut in two by cutting a single Green's function line.
An approximate self-energy will sum over a subset of these diagrams, which
can be specified in terms of the diagram skeletons ($S$), and the Green's functions $\mathbf{G}_\Sigma(\omega)$
and interactions $V$ within them. We  thus denote a self-energy approximation by $\boldsymbol{\Sigma}^S[\mathbf{G}_\Sigma, V]$.
A {\it self-consistent} approximate self-energy is one where the Green's function used in the self-energy
diagrams satisfies the Dyson equation, i.e. $\mathbf{G} = \mathbf{G}_\Sigma$.

In Green's function embedding, the fragment Green's function $\mathbf{G}_A (\omega)$, is adjusted through
another self-energy $\boldsymbol{\Delta}_A(\omega)$, dual to $\mathbf{G}_A(\omega)$. To prevent confusion with the self-energy arising due to interactions in Eq.~(12), we term $\boldsymbol{\Delta}_A(\omega)$ the hybridization. Together with the self-energy approximation $\boldsymbol{\Sigma}_A(\omega)$, this gives the fragment
Dyson equation
\begin{align}
\mathbf{G}_A(\omega) = \mathbf{g}_A(\omega) + \mathbf{G}_A(\omega) (\boldsymbol{\Delta}_A(\omega) + \boldsymbol{\Sigma}_A(\omega)) \mathbf{g}_A(\omega),
\label{eq:embed_dyson}
\end{align}
where $\mathbf{{g}}_A(\omega)$ is the non-interacting Green's function of the system computed {\it in isolation}, i.e. with no couplings to the environment or any interactions. {Inverting Eq.~(13) gives the hybridization as
\begin{align}
  \boldsymbol{\Delta}_A(\omega) = \mathbf{g}_A(\omega)^{-1} - \mathbf{G}_A(\omega)^{-1} - \boldsymbol{\Sigma}_A(\omega) \label{eq:hyb}.
  \end{align}
}
$\boldsymbol{\Delta}_A(\omega)$ is the analog of the DFT embedding potential $v_A$, and Eq.~(14)
is analogous to Eq.~(6).
Similarly to $v_A$, $\boldsymbol{\Delta}_A(\omega)$ contains effects from electron delocalization into the environment
as well as environment Coulomb interactions. However, unlike in DFT embedding,
the inversion from $\mathbf{G}(\omega) \to \boldsymbol{\Delta}(\omega)$ is explicit through Eq.~(14), and no
iterative technique is necessary.

We now discuss  two different contexts in which Green's function embedding is applied.
These correspond to different self-energy approximations for $A$ and the environment $B$, and
parallel the contexts appearing in DFT embedding.
The simplest is to describe both $A$ and the full system at the mean-field level.
Then, $\boldsymbol{\Sigma}_A(\omega)$ and $\boldsymbol{\Sigma}(\omega)$ 
correspond to the respective mean-field Coulomb and exchange terms (and thus have no frequency dependence).
Similarly to DFT in DFT embedding, savings result when we use different levels of mean-field
for $A$ and the full-system, for example by forgoing self-consistency in the environment.
A common application of the latter is to impurities in crystals, where $\mathbf{G}(\omega)$ 
is first computed using translational invariance in the periodic crystal, and relaxation of the environment
is ignored when the impurity is introduced~\cite{grimley1974chemisorption}. In
molecular junctions, $\boldsymbol{\Delta}_A(\omega)$ is similarly obtained for
a semi-infinite electrode, and assumed to be unchanged on the introduction of the bridging molecule~\cite{brandbyge2002density}.

A second context is to perform Green's function embedding with more sophisticated approximations for $\boldsymbol{\Sigma}_A(\omega)$ and
$\boldsymbol{\Sigma}(\omega)$. This is similar to wavefunction in DFT embedding, as there are now
two different correlation treatments which must be bridged.
We denote the ``high-level'' approximation $S_H$ and the ``low-level'' approximation $S_L$. The low-level
approximation is used to construct a self-energy including all interactions in the
full problem, $\boldsymbol{\Sigma}^{S_L}[\mathbf{G}, V](\omega)$, and the high-level approximation is used to construct a self-energy considering
interactions only in the subsystem $A$, $\boldsymbol{\Sigma}^{S_H}_A[\mathbf{G}_A, V_A](\omega)$. The
composite self-energy for the full problem is
\begin{align}
  \boldsymbol{\Sigma}(\omega) = \boldsymbol{\Sigma}^{S_L}[ \mathbf{G}, V] + \boldsymbol{\Sigma}_A^{S_H}[ \mathbf{G}_A, V_A](\omega) - 
\boldsymbol{\Sigma}^{S_L}_A[ \mathbf{G}_A, V_A](\omega) \label{eq:twolevelselfenergy}
\end{align}
where $\boldsymbol{\Sigma}^{S_L}_A[ \mathbf{G}_A, V_A]$ indicates the part of $\boldsymbol{\Sigma}^{S_L}[ \mathbf{G}, V]$
involving only Green's functions and interactions within fragment $A$. Eq.~(15) 
is analogous to the expression for
the exchange-correlation potential Eq.~(9) in wavefunction in DFT embedding. In fact, one common way to
compute $\boldsymbol{\Sigma}^{S_H}_A[\mathbf{G}_A, V_A](\omega)$ is by carrying out a wavefunction calculation on the subsystem $A$ in the presence of additional fictitious ``bath'' orbitals that reproduce the effects of the hybridization $\boldsymbol{\Delta}_A$~\cite{Chan2011a}. The subsystem
plus  bath orbitals is known as a quantum impurity problem. We return to impurity problems
in the context of density matrix embedding theory. Self-consistency of the Green's functions is  obtained by solving Eq.~(14) and Eq.~(15)  for the hybridization and self-energy.

Dynamical mean-field theory~\cite{georges1992hubbard,Kotliar2006,Chan2011a} (DMFT) (which here refers to both single-site and cluster extensions)
provides a widely used example of a higher-level
self-energy embedding within a lower-level treatment. 
In DMFT, we divide the full problem of interest (commonly a crystal) into multiple fragments $A$
containing strongly correlated orbitals (typically transition metal $d$ and $f$ orbitals)
for which a high-level self-energy approximation in each fragment, $\boldsymbol{\Sigma}_A(\omega)$, is computed.
The composite self-energy in Eq.~(15) becomes
\begin{align}
  \boldsymbol{\Sigma}(\omega) = \boldsymbol{\Sigma}^{S_L}( \mathbf{G}, V) + \sum_A [\boldsymbol{\Sigma}^{S_H}_A( \mathbf{G}_A, V_A) - \boldsymbol{\Sigma}^{S_L}_A(\mathbf{G}_A, V_A)] \label{eq:dmftselfenergy}
\end{align}
where the sum over $A$ reflects a summation over the fragments. For a crystal, the self-energy
in each cell is identical. Self-consistency of Eq.~(16) with
 $\boldsymbol{\Delta}_A(\omega)$ for each cell $A$ yields the cellular DMFT equations~\cite{Biroli2002}.

The low-level approximation in DMFT is often chosen to be a DFT treatment.
This combination is called DFT+DMFT~\cite{Kotliar2006}, and has been widely applied to correlated materials,
especially to compute the density-of-states observed in photoemission experiments~\cite{Kotliar2006}. It has also been
used, less commonly, in molecular applications,
for example, to obtain correlation corrections to the conductance of a molecular junction~\cite{Jacob2010}, and
qualitative features of the electronic structure in transition metal clusters and complexes~\cite{Turkowski2012}. 

Unfortunately, the combination of DFT with DMFT suffers from similar problems
to wavefunction in DFT embedding, such as partial double counting of interactions, and this has been
a barrier to chemical accuracy. Combining DMFT 
with a low-level Hartree-Fock self energy avoids this issue, as the diagrams can be
correctly subtracted in Eq.~(16), and has recently been explored in
small molecules~\cite{Lin2011}, and in a minimal basis cubic hydrogen solid~\cite{Chan2011a}.
Unfortunately,
this treatment provides no description of correlations outside of the fragments, and
 is thus also not quantitative. Incorporating
``non-local'' correlations into a DMFT description is a topic of current research,
and strategies include 
using a random-phase approximation (RPA) for the non-local correlations, which modifies both the low-level
self-energy, as well as screens the Coulomb interaction $V_A \to V_A^{RPA}(\omega)$ appearing in the high-level
self-energy approximation; and using a pure self-energy approximation, such as
the self-consistent second-order self-energy~\cite{nooijen1995second}. The latter has been 
explored by Zgid and coworkers~\cite{Phillips2014,Lan2015}. 

The strengths of Green's function embedding are the
analytic expressions for the energy and hybridization, and the diagrammatic interpretation
of the self-energy approximations. However, fully ab-initio applications lag behind
those of DFT embedding, because ab-initio quantum chemistry
methods are primarily developed for single states (such as the ground-state) rather than the Green's function,
and  computing time-dependent Green's functions is more expensive than
computing time-independent observables.

\section{Density matrix embedding}
  
The practical complexity of working with Green's functions motivates
a third formulation of embedding: density matrix embedding, where the quantity of interest is the
single-particle density matrix, $\boldsymbol{\gamma}$, where $\gamma_{ij} = \langle \Psi | a^\dag_i a_j | \Psi\rangle$.
It appears straightforward to formulate embedding with the density matrix, as it
 interpolates in complexity between the density and the Green's function. Indeed, one can
formally define a density matrix embedding to parallel DFT (and Green's function) embedding, starting from
the energy functional of the density matrix, and using stationarity to define an embedding operator $\mathbf{v}_A$,
\begin{align}
  \mathbf{v}_A = \frac{\delta \Delta E}{\delta \boldsymbol{\gamma}_A} \label{eq:embed_1op}
  \end{align}
which, when substituted into the fragment Euler equation, yields the exact fragment density matrix $\boldsymbol{\gamma}_A$.

Unfortunately, there is a complication that is unique to the density matrix formulation.
In  DFT embedding, $\Delta E$ and  the embedding potential are defined with respect
to a closed non-interacting system, and the latter
acts to adjust the reference density to
match the fragment density. (A similar statement may be made with
respect to the Green's function for Green's function embedding). However,
the {\it density matrix} of a closed non-interacting
system must be idempotent, while the density matrix of a general (open) fragment need not be, and
the two in general cannot be matched via an embedding operator. 
Instead, one  is forced to consider  more complex interacting reference systems (to represent non-idempotent fragment density matrices)
or more complicated interacting density matrix functionals that minimize
to non-idempotent solutions. Some of the more complex reference systems  considered include ones where the ground-state is modeled
by a geminal wavefunction~\cite{Tsuchimochi2015,Pernal2016}, as well an impurity-like
formulation, where the reference is the size of the full system, but with interactions restricted only
to the fragment of interest~\cite{Senjean2016}.

The above complication reflects the essential physical difference between
an open system, which is entangled with its environment and should be
described by a mixed state, and a closed system, described by a pure state.
Rather than using a more complicated pure-state to mimic a non-idempotent density matrix, one can instead model
the open fragment as part of a closed system, by introducing additional bath degrees of freedom.
This is the same physical idea used in the impurity representation
of the hybridization in Green's function embedding. 
 Of course, if the bath has the same complexity as the rest of the system, nothing is gained,
but in practice, the size of the bath can be  significantly less than the size of the environment~\cite{Sun2014,Wouters2016}.
A simple example is the link orbitals in QM/MM calculations, 
a small set of extra atomic orbitals chosen to saturate the dangling bonds of the QM fragment.
Other embedding approaches, such as divide and conquer, similarly introduce buffer orbitals~\cite{Yang1995}, using
orbitals  spatially close to the fragment region. 

The choice of additional bath degrees of freedom appears to have
some arbitrariness, but a construction that is provably optimal at the mean-field level  has recently been provided
 by the DMET (density matrix embedding theory) of Knizia and Chan~\cite{Knizia2012,Knizia2013}.
A key property of the DMET bath is that it is (at most)
 the same size as the fragment, and thus rigorously removes most of the degrees of freedom in the environment.
DMET thus avoids the much larger baths associated with Green's function embeddings~\cite{Knizia2012}.
To see that the environment can be compressed to this size, consider 
the case where one has the exact wavefunction $\ket{\Psi}$ for
the full problem. This wavefunction
may be rewritten (via the Schmidt decomposition) in terms of states that live
solely in the fragment, and environment Hilbert spaces, $\{ \ket {\alpha_i} \}$, $\{ \ket {\beta_i} \}$, respectively
\begin{align}
\ket{\Psi}=\sum_i^D \lambda_i \ket{{\alpha}_i}\ket{{\beta}_i}
\end{align}
The summation is over $D$ terms, the dimension of the fragment Hilbert space, and
the decomposition defines $D$ exact bath states $\ket{\beta_i}$. The exact
wavefunction can then be expressed within the reduced Hilbert space $\{ \ket{\alpha_i} \} \otimes \{ \ket{\beta_j} \}$.
When  $\ket{\Psi}$ is a Slater determinant (e.g. the ground-state of a mean-field Hamiltonian {$\hat{f}$}) then the Schmidt decomposition takes a particularly simple form. In particular, if the fragment has $d$ orbitals, then the bath states
are spanned exactly by the Hilbert space of $d$ (partially occupied) bath orbitals, with all other
orbitals  either completely filled (core orbitals) or empty. The partially filled bath orbitals
are the eigenvectors (with partial occupancy) of the environment block of the mean-field density matrix.
This fragment plus bath representation of a mean-field wavefunction
becomes the reference system in DMET, a non-interacting problem of twice the size of the fragment.
Projected into this representation,
the  mean-field ground-state of {$\hat{f}$} can reproduce any fragment density matrix by
 augmenting with a suitable fragment operator, $\hat{f} \to \hat{f}+\hat{v}_A$.
 Since the bath orbitals capture the effects of embedding at the mean-field level,
 $\hat{v}_A$ serves to encode additional correlation effects beyond the mean-field treatment, and is thus analogous
to the exchange-correlation potential in DFT, or self-energy in Green's function embedding.

Mean-field in mean-field embedding in DMET can be formulated using different levels of mean-field theory
to define  bath orbitals from the full problem, and to model the smaller fragment plus bath representation.
However, DMET has so far mainly been applied using a correlated wavefunction description
of the fragment plus bath, on top of a mean-field reference. Denoting the
 mean-field description by $S_L$, 
the DMET bath orbitals are obtained from the mean-field reference $\Psi^{S_L}$ (the ground-state of $\hat{f} + \hat{v}_A$)
 which defines a fragment density matrix $\boldsymbol{\gamma}_A$.
 The correlated energy of the {\it full} problem is then
\begin{align}
  E^{S_H}[\boldsymbol{\gamma}_A] = \min_{\Psi_A^{S_H}}\langle \Psi_A^{S_H} | \hat{H} | \Psi_A^{S_H} \rangle \label{eq:dmint}
   \end{align}
where the correlated wavefunctions are defined in the DMET ``active'' space
of fragment $A$ plus its bath and core orbitals. The functional dependence on
 $\boldsymbol{\gamma}_A$ enters in Eq.~(19) through the definition of the bath,
but  $\boldsymbol{\gamma}_A$ is in general different from the high level density matrix $\boldsymbol{\gamma}_A^{S_H}$,
obtained from $\Psi_A^{S_H}$.
Self-consistency adjusts $\hat{v}_A$ such that $\boldsymbol{\gamma}_A =\boldsymbol{\gamma}^{S_H}_A$. The numerical procedure
to do so involves a non-interacting inversion $\boldsymbol{\gamma}_A  \to \mathbf{v}_A$, and
is  analogous to the inversion $\rho_A \to v_A$ in DFT embedding, with related issues of representability~\cite{Tsuchimochi2015,Wouters2016}.

Although $E^{S_H}[\boldsymbol{\gamma_A}]$ is an energy for the full problem,
correlations are only included close to fragment $A$, as the bath orbitals are typically
localized close to $A$. This is acceptable for intensive quantities (such as local
reactions or excitations). However, for an extensive correlated treatment (as desired in a condensed phase
problem) one must embed with multiple fragments.
Then, each fragment $A$ yields a separate high-level wavefunction for the full problem, $\Psi_A^{S_H}$,
and expectation values must be assembled from the different fragment wavefunctions.
It can be expected that
expectation values for a given $\Psi_A^{S_H}$ are most accurate close to fragment $A$,
and this is reflected in the partitioning of the contributions.
For example, to compute the density matrix element $\gamma_{ij} = \langle a^\dag_i a_j\rangle$, we define
\begin{align}
  \gamma_{ij} =
  \begin{cases}
    \langle \Psi_A^{S_H} | a^\dag_i a_j | \Psi_A^{S_H} \rangle &  i, j \in A \\
    \frac{1}{2} (\langle \Psi_A^{S_H} | a^\dag_i a_j | \Psi_A^{S_H} \rangle + \langle \Psi_{A'}^{S_H} | a^\dag_i a_j | \Psi_{A'}^{S_H} \rangle & i \in A, j \in A'.
  \end{cases}
\end{align}
Analogous partitionings can be defined for more complicated expectation values. 

Because its initial development was motivated by DMFT, the majority of DMET applications  have been to
correlated lattice models~\cite{bulik2014density,Tsuchimochi2015,Zheng2016}, where the computational simplicity of the method
has enabled very accurate results to be achieved using large fragments, for example, for the 2D Hubbard model~\cite{Zheng2016}.
Importantly,  self-consistency in DMET, much like in DMFT, allows for non-trivial phases, such
as superconductivity in repulsive systems~\cite{Zheng2016}.
Related to the physics of such lattice models are problems of strong correlation in chemical settings.
The DMET bath allows the  formalism to accurately
treat fragments when they are bonded to their environment, even
when such a bond is stretched or dissociated, as has been demonstrated through
accurate calculations of the dissociation curves of molecular chains and rings\cite{Knizia2013,Wouters2016}. 
An exciting area of  application is to  reduce the cost of high-level correlated wavefunction calculations in
solids, particularly for small-band gap systems such as metals. Here, DMET is used to treat a unit-cell (or a small
set of them) while the extra bath orbitals  ameliorate
the finite size effects associated with the small cell. Demonstrations on crystals in 1, 2, and 3 dimensions by Scuseria and coworkers
have shown promise~\cite{bulik2014electron}. 
Recent work has focused on extensions of DMET, for example, to spectra\cite{Booth2015}, where
the additional bath orbitals are modified to carry a time-dependence that reproduces
the linear response of a mean-field wavefunction. This provides a rigorous way to embed
excited states, which are often much more delocalized than the ground-state, and thus
 more sensitive to the open nature of a chemical fragment.

\section{Conclusions}

Quantum embedding is a natural computational framework
in which to think about complex systems.  We have focused on three 
embedding approaches based on the single-particle density, Green's function, and density matrix
respectively. While we have only been able to give a short description of these approaches,
we have tried to bring out their common intellectual structure.

There remain many frontier methodological areas; for example,
excited states and dynamics in density functional embedding, and  more efficient ab-initio technology in Green's function and density matrix embedding.
New  application areas are emerging, for example, in  biomolecular and condensed phase simulations. In some
cases it is necessary to  include classical
embeddings, such as through QM/MM  as well. While we cannot predict the future development of the field, the growing activity strongly suggests that quantum embedding methods will remain a key part of simulating complex systems
for many years to come.

\section{Biographical information}

Qiming Sun received his Ph. D. from Peking University. After postdoctoral
work at Princeton University, he joined the California Institute of Technology as a staff scientist. He is the principal developer of the PySCF quantum chemistry package.

Garnet Kin-Lic Chan received his Ph. D. from the University of Cambridge, and
carried out postdoctoral work as a Junior Research Fellow
at Christ's College, Cambridge, and as a Miller Research Fellow
at the University of California, Berkeley. After
appointments at Cornell University and Princeton University, he joined the California Institute of Technology, where he is currently the
Bren Professor of Chemistry. 

\section{Acknowledgments}

G. K.-L. Chan acknowledges support from the US National Science Foundation through grant no. NSF-CHE-1265277.

\bibliography{embed}

\end{document}